\documentclass{sig-alternate}

\usepackage[english]{babel}
\usepackage{times}
\usepackage{amsmath, amssymb}
\usepackage{algorithm}
\usepackage{listings}
\usepackage{varioref}
\usepackage{subfigure}
\usepackage{color}
\usepackage{xspace}
\usepackage{tikz}
\usepackage{varioref}
\usepackage[numbers,sort&compress]{natbib}
\usetikzlibrary{shapes}

\usepackage{url}

\floatname{algorithm}{Listing}

\newcommand\jbotsim{{\textsc{JBotSim}}\xspace}
\renewcommand\tt{\ttfamily \small}

\newcommand\cmd[1]{{\tt \footnotesize #1}\xspace}

\title{JBotSim: a Tool for Fast Prototyping of Distributed Algorithms in Dynamic Networks\thanks{A shorter version appeared in SIMUTOOLS 2015. For up to date information, the reader is advised to visit {\tt https://jbotsim.io}.}}
\author{Arnaud Casteigts \medskip\\LaBRI, University of Bordeaux\medskip\\{\tt \large arnaud.casteigts@labri.fr} \and
Remi Laplace \medskip\\LaBRI, University of Bordeaux\medskip\\{\tt \large remi.laplace@labri.fr}}

\definecolor{darkdarkgray}{gray}{0.20}
\begin{document}
\maketitle

\begin{abstract}
\jbotsim is a java library that offers basic primitives for prototyping, running, and visualizing distributed algorithms in dynamic networks. With \jbotsim, one can implement an idea in minutes and interact with it ({\it e.g.}, add, move, or delete nodes) while it is running. \jbotsim is well suited to prepare live demonstrations of algorithms to colleagues or students; it can also be used to evaluate performance at the algorithmic level (number of messages, number of rounds, etc.). Unlike most simulation tools, \jbotsim is not an integrated environment. It is a lightweight library to be used in java programs. In this paper, we present an overview of its distinctive features and architecture.
\end{abstract}

\section{Introduction}
JBotSim is an open source simulation library that is dedicated to distributed algorithms in dynamic networks. It was developed with the purpose in mind to make it possible to implement an algorithmic idea in minutes and interact with it while it is running ({\it e.g.}, add, move, or delete nodes). Besides interaction, \jbotsim can also be used to prepare live demos of an algorithm and to show it to colleagues or students, as well as to assess the algorithm performance. \jbotsim is not a competitor of mainstream simulators such as NS3~\cite{ns3}, OMNet~\cite{omnet}, or The One~\cite{theone}, in the sense that it does not aim to implement real-world networking protocols. Quite the opposite, \jbotsim aims to remain technology-insensitive and to be used at the algorithmic level, in a way closer in spirit to the ViSiDiA project (a general-purpose platform for distributed algorithms). Unlike ViSiDiA, however, \jbotsim natively supports mobility and dynamic networks (as well as wireless communication). Another major difference with the above tools is that it is a {\em library} rather than a software: its purpose is to be used in other programs, whether these programs are simple scenarios of full-fledged software. Finally, \jbotsim is distributed under the terms of the LGPL licence, which makes it easily extensible by the community.

Whether the algorithms are centralized or distributed, the natural way of programming in \jbotsim is event-driven: algorithms are specified as subroutines to be executed when particular events occur (appearance or disappearance of a link, arrival of a message, clock pulse, {\it etc.}). 
Movements of the nodes can be controlled either by program or by means of live interaction with the mouse (adding, deleting, or moving nodes around with left-click, right-click, or drag and drop, respectively). These movements are typically performed while the algorithm is running, in order to visualize it or test its behavior in challenging configurations. 

The present document offers a broad view of \jbotsim's main features and design traits. We start with preliminary information in Section~\ref{sec:preliminaries} regarding installation and documentation. Section~\ref{sec:overview} reviews \jbotsim's main components and specificities such as programming paradigms, clock scheduling, user interaction, or global architecture. Section~\ref{sec:zoom} zooms on key features such as the exchange of messages between nodes, graph-level APIs, or the creation of online demos. Finally, we discuss in Section~\ref{sec:extensions} some extensions of \jbotsim, including Ti{\it k}Z exportation feature and edge-markovian dynamic graph generator.

Besides its features, the main asset of JBotSim is its simplicity of use -- an aim that was pursued at the cost of re-writing it several times from scratch (the API is now in version 1.0). 

\section{Practical aspects}
\label{sec:preliminaries}
In this short section, we show how to run \jbotsim with a first basic example (this step is not required to keep reading the present document). We also provide links to online documentation and examples, for readers who would like to explore \jbotsim's features beyond this paper.

\subsection{Fetching {\sc \large JBotSim}}

Since version 1.0.0, \jbotsim is available on {\it Maven}. The natural way to retrieve it (along with the {\it Javadoc} and source code) by using its {\it Maven coordinates} (see Listing~\ref{algo:mavencoordinates}), in your project.

\begin{algorithm}[H]
\begin{lstlisting}
<dependency>
  <groupId>io.jbotsim</groupId>
  <artifactId>jbotsim-all</artifactId>
  <version>1.x.x</version> // replace with current version
</dependency>
\end{lstlisting}
\caption{\label{algo:mavencoordinates} {\it Maven coordinates} for \jbotsim's latest stable version}
\end{algorithm}

For more specific cases, you may want to retrieve only some of \jbotsim's artifacts (e.g. only the core without UI). It is also possible to use a standalone {\em jar} package as it used to be before version 1.0.0. For more explanation, please refer to \jbotsim's website~\cite{jbotsim-website} or the related documentation on the GitHub repository~\cite{jbotsim-github}.

\subsection{First steps}
As a first program, one can copy the code from Listing~\ref{algo:helloworld} into a file named \cmd{HelloWorld.java}.

\begin{algorithm}[H]
\begin{lstlisting}[language=Java]
import io.jbotsim.core.Topology;
import io.jbotsim.ui.JViewer;

public class HelloWorld{
    public static void main(String[] args){
        Topology tp = new Topology();
        new JViewer(tp);
        tp.start();
    }
}
\end{lstlisting}
\caption{\label{algo:helloworld} HelloWorld with \jbotsim}
\end{algorithm}


By running the program, one should see an empty gray surface in which nodes can be added, moved, or deleted using the mouse.
Please note that this {\it HelloWorld} is more developped on both the \jbotsim's website~\cite{jbotsim-website} and the GitHub repository~\cite{jbotsim-github}.

\subsection{Sources of documentation}
\label{sec:documentation}

In this document, we provide a general overview of what \jbotsim is and how it is designed. This is by no means a comprehensive programming manual. The reader who wants to explore further some features or develop complex programs with \jbotsim is referred to the API documentation (see {\em javadoc} on the website~\cite{jbotsim-website} and in your IDE).

Examples can also be found on \jbotsim's website, together with comments and explanations. These examples offer a good starting point to learn specific components of the API from an {\em operational} standpoint --~the present document essentially focuses on {\em concepts}. Most of the online examples feature embedded videos. Finally, most of the examples in this paper are also available on \jbotsim's website. Feel free to check them when the code given here is incomplete (e.g. we often omit package imports and \cmd{main()} methods for conciseness).

\section{Features and architecture}
\label{sec:overview}
This section provides an overview of \jbotsim's key features and discusses the reason why some design choices were made. We review topics as varied as programming paradigms, clock scheduling, user interaction, and global architecture.

\subsection{Basic features of nodes and links}

\begin{figure}[h]
  \centering
  \includegraphics[width=8cm]{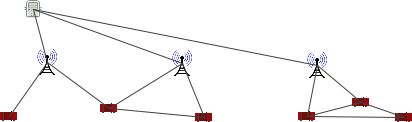}
  \caption{\label{fig:highway} A highway scenario composed of vehicles, road-side units, and central servers. Part of the network is {\it ad hoc} (and wireless); the rest is infrastructured (and wired).}
\end{figure}

\jbotsim consists of a small number of classes, the most central being \cmd{Node}, \cmd{Link}, and \cmd{Topology}. The contexts in which dynamic networks apply are varied. In order to accommodate a majority of cases, these classes offer a number of conceptual variations around the notions of nodes and links. Nodes may or may not possess wireless communication capabilities, sensing abilities, or self-mobility. They may differ in clock frequency, color, communication range, or any other user-defined property. Links between the nodes account for potential communication among them. The nature of links varies as well; a link can be directed or undirected, as well as it can be wired or wireless -- in the latter case \jbotsim's topology will update the set of links {\em automatically}, as a function of nodes distances and communication ranges.

\begin{figure}[h]
  \centering
  \includegraphics[width=7cm]{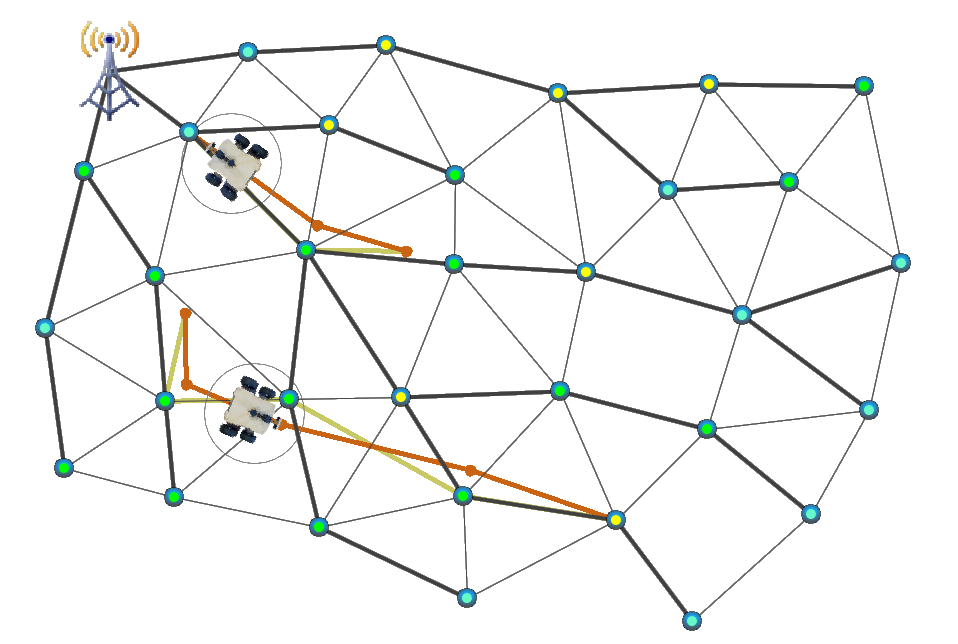}
  \caption{\label{fig:recharge} A data collection scenario, whereby ground robots recharge a wireless sensor network.}
\end{figure}

Figures~\ref{fig:highway} and~\ref{fig:recharge} illustrate two different contexts. Figure~\ref{fig:highway} depicts a highway scenario where three types of nodes are used: vehicles, road-side units (towers), and central servers. This scenario is semi-infrastructured: Servers share a dedicated link with each tower. These links are {\em wired} and thus exist irrespective of distance. On the other hand, towers and vehicles communicates through wireless links that are automatically updated.

Figure~\ref{fig:recharge} illustrates a purely {\it ad hoc} scenario in which a {\em sink} node (depicted by the radio tower) collects data from a wireless sensor network.
In order expand battery life, measurements are routed along a dynamically maintained tree rooted at the {\em sink}.
On top of this, ground robots are tasked to periodically recharge the sensors. 
For simplicity, a robot fully recharges a sensor whenever it is within its {\em sensing range} (depicted by a surrounding circle). 
In the presented example, once a robot has randomly chosen a sensor destination, it sets up a trajectory trying and maximize the number of sensors recharged during its journey (depicted in orange).

Besides nodes and links, the concept of topology is central in \jbotsim. Topologies can be thought of as {\it containers} for nodes and links, together with dedicated operations like updating wireless links. They also play a central role in \jbotsim's event architecture, as explained later on.

\subsection{Distributed vs. centralized algorithms}
\label{sec:distributed}
\jbotsim supports the manipulation of centralized or distributed algorithms (possibly simultaneously). The natural way to implement a distributed algorithm is by extending the \cmd{Node} class, in which the desired behavior is implemented. Centralized algorithms are not constrained to a particular model, they can take the form of any standard java class. 

\paragraph{Distributed algorithm} \jbotsim comes with a default type of node that is implemented in the \cmd{Node} class. This class provides the most general features a node could have, including primitives for moving, exchanging messages, or tuning basic parameters (e.g. communication range and sensing range). Distributed algorithms are naturally implemented through adding specific features to this class. Listing~\ref{algo:movingnode} provides a basic example in which the nodes are endowed with self-mobility.
\begin{algorithm}[h]
\begin{lstlisting}[language=Java]
import io.jbotsim.core.Node;

public class MovingNode extends Node{
    public MovingNode(){
        setDirection(Math.random() * 2*Math.PI);
    }
    public void onClock(){
        move(2);
    }
}
\end{lstlisting}
\caption{\label{algo:movingnode}Extending the \cmd{Node} class}
\end{algorithm}
The class relies on a key mechanism in \jbotsim: performing periodic operations that are triggered by the pulse of the system clock. This is done by overriding the \cmd{onClock()} method, which is called periodically by JBotSim's engine (by default, at every pulse of the clock).
The rest of the code is responsible for moving the node, setting a random direction at construction time (in radian), then moving in this direction periodically. (More details about the movement API can be found online.)

Once this class is defined, new nodes of this type can be added to the topology in the same way as if they were of class Node, as illustrated in Listing~\ref{algo:addNode}.  
\begin{algorithm}[h]
\begin{lstlisting}[language=Java, firstline=5, lastline=11]
public static void main(String[] args){
    Topology tp = new Topology(400,300);
    for (int i=0; i<10; i++)
        tp.addNode(-1,-1, new MovingNode());
    new JViewer(tp);
    tp.start();
}
\end{lstlisting}
\caption{\label{algo:addNode} Adding nodes manually}
\end{algorithm}
Here, a new topology is created first, then 10 nodes of the desired type (here, \cmd{MovingNode}) are added through the \cmd{addNode()} method at random locations (using $-1$ for $x$-coordinate or $y$-coordinate generates a random location for that coordinate; here both are random). 
In order to use \cmd{MovingNode} through the GUI, for instance when clicking with the mouse to add new nodes, this class must be registered as a {\em node model}. This is done by calling method \cmd{setDefaultNodeModel()} with the class itself as an argument, as shown on Listing~\ref{algo:setModel}. \jbotsim will create new instances on the fly, using reflexivity. Several models can be registered simultaneously, using \cmd{setNodeModel()} with an additional argument that corresponds to the model name. If several models exist, \jbotsim's GUI (the {\em viewer}) displays a selection list when a node is added by the user (in this example, only one model is set and it is used by default). 
\begin{algorithm}[h]
\begin{lstlisting}[language=Java, firstline=5, lastline=10]
public static void main(String[] args){
    Topology tp = new Topology(400, 300);
    tp.setDefaultNodeModel(MovingNode.class);
    new JViewer(tp);
    tp.start();
}
\end{lstlisting}
\caption{\label{algo:setModel} Using a defined node as default}
\end{algorithm}
In the scenario of Figure~\ref{fig:highway}, however, left-clicking on the surface would give the choice between {\em car}, {\em tower}, and {\em server}, the names of the three registered models for that scenario.

\paragraph{Centralized algorithms}

There are many reasons why a centralized algorithm can be preferred over a distributed one. The object of study might be centralized in itself (e.g. network optimization, scheduling, graph algorithms in general). It may also be simpler to simulate distributed things in a centralized way. Listing~\ref{algo:globalRWP} implements such a version of the {\em random waypoint} mobility model, in which nodes repeatedly move toward a randomly selected destination, called {\em target}. 
\begin{algorithm}[h]
\begin{lstlisting}[language=Java, firstline=7]
public class CentralizedRWP implements ClockListener {
  Topology tp;
  public CentralizedRWP (Topology tp){
     this.tp = tp;
     tp.addClockListener(this);
  }
  public void onClock(){
     for (Node n : tp.getNodes()){
        Point target=(Point)n.getProperty("target");
        if (target == null || n.distance(target)<1){
           target = new Point(Math.random()*400,
                                                Math.random()*300);
           n.setProperty("target", target);
           n.setDirection(target);
        }
        n.move();
     }
  }
  public static void main(String[] args){
     Topology tp = new Topology(400,300);
     new CentralizedRWP(tp);
     new JViewer(tp);
     tp.start();
  }
}
\end{lstlisting}
\caption{\label{algo:globalRWP} Centralized version of Random Waypoint}
\end{algorithm}
Unlike a distributed implementation, the movements of nodes are here driven by a {\em global} loop at every pulse of the clock. For each node, a target is created if it does not exists yet or if it has just been reached; then the node's direction is set accordingly and the node is moved (by default, by 1 unit of distance). For convenience, the \cmd{main()} method is included in the same class.

Notice the use of \cmd{setProperty()} and \cmd{getProperty()} in this example. These methods allow to store any object directly into a node, using a key/value scheme (where key is a string). Both \cmd{Link} and \cmd{Topology} objects offer the same feature.

\subsection{Architecture of the event system}

\begin{figure*}
  \centering
  \tikzset{class/.style={fill=none,rectangle, inner sep=3pt}}
\tikzset{interface/.style={fill=none,rectangle,rounded corners=5pt, inner sep=3pt, font=\footnotesize}}
\tikzset{cx/.style={inner sep=0pt}}
\begin{tikzpicture}[yscale=1.2, xscale=.9]
  \path (0,1) node[class] (Link){Link};
  \path (0,2) node[class] (Node){Node};
  \path (0,3) node[class] (Topology){Topology};
  \path (-2,1.2) coordinate (cxMessageListener){};
  \path (-2,3.8) coordinate (cxMovementListener){};
  \path (-2,2.5) coordinate (cxConnectivityListener){};
  \path (2,2.5) coordinate (cxTopologyListener){};
  \path (2,1.2) coordinate (cxPropertyListener){};
  \path (2,3.8) coordinate (cxClockListener){};
  \path (cxPropertyListener)+(3,0) node[interface,right] (PropertyListener){PropertyListener};
  \path (cxMovementListener)+(-3,0) node[interface,left] (MovementListener){MovementListener};
  \path (cxConnectivityListener)+(-3,0) node[interface,left] (ConnectivityListener){ConnectivityListener};
  \path (cxTopologyListener)+(3,0) node[interface,right] (TopologyListener){TopologyListener};
  \path (cxMessageListener)+(-3,0) node[interface,left] (MessageListener){MessageListener};
  \path (cxClockListener)+(3,0) node[interface,right] (ClockListener){ClockListener};

  \draw (Topology.east)--(cxTopologyListener);
  \draw (Link.east)--(cxPropertyListener);
  \draw (Node.east)--(cxPropertyListener);
  \draw (Topology.east)--(cxPropertyListener);
  \draw (Topology.west)+(.1,0) -- (cxMessageListener);
  \draw (Topology.west)+(.1,0) -- (cxMovementListener);
  \draw (Topology.west)+(.1,0)--(cxConnectivityListener);
  \draw (Topology.east)--(cxClockListener);

  \tikzstyle{every path}=[shorten >=4pt]
  \tikzstyle{every node}=[font=\scriptsize, inner sep=1.5pt,midway]
  \path[->] (cxClockListener) edge node[below]{onClock()} (ClockListener.west);

  \path[bend left=10, ->] (cxTopologyListener) edge node[above,inner sep=1pt]{onNodeAdded()} ([yshift=1pt]TopologyListener.west);
  \path[bend right=10, ->] (cxTopologyListener) edge node[below]{onNodeRemoved()} ([yshift=-1pt]TopologyListener.west);

  \path[->] (cxPropertyListener) edge node[below]{onPropertyChanged()} (PropertyListener.west);

  \path[->] (cxMessageListener) edge node[below]{onMessage()} (MessageListener.east);

  \path[->] (cxMovementListener) edge node[below]{onMovement()} (MovementListener.east);

  \path[bend right=10, ->] (cxConnectivityListener) edge node[above,inner sep=1pt]{onLinkAdded()} ([yshift=1pt]ConnectivityListener.east);
  \path[bend left=10, ->] (cxConnectivityListener) edge node[below]{onLinkRemoved()} ([yshift=-1pt]ConnectivityListener.east);

  \tikzstyle{every node}=[]
\end{tikzpicture}
\caption{\label{fig:events}Main sources of events and corresponding interfaces in \jbotsim.}
\end{figure*}
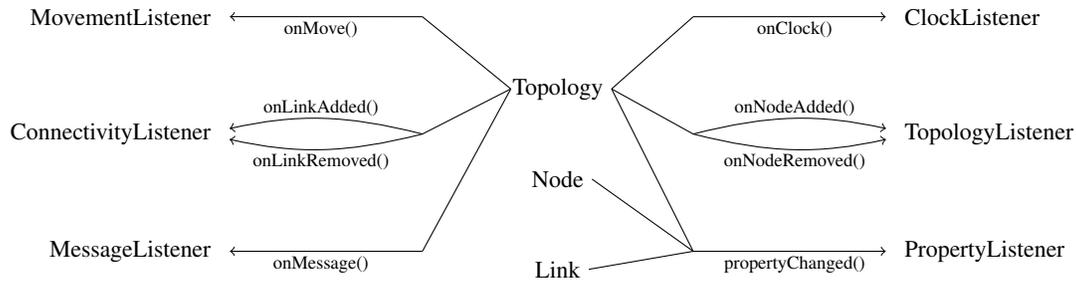

So far, we have seen one type of event: clock pulses, to be listened to through the \cmd{ClockListener} interface. \jbotsim offers a number of such events and interfaces, some of which are ubiquitous. The main ones are depicted on Figure~\vref{fig:events}. This architecture allows one to specify dedicated operations in reaction to various events. For instance, one may ask to be notified whenever a link appears or disappears somewhere. Same for messages, which are typically listened to by the nodes themselves or can be watched at a global scale (e.g. to keep a log of all communications). In fact, every node is automatically notified for its own events; it just needs to override the corresponding methods from the parent class \cmd{Node} in order to specify event handlings (e.g. \cmd{onClock()}, \cmd{onMessage()}, \cmd{onLinkAdded()}, \cmd{onSensingIn()}, \cmd{onSelection()}, {\it etc.}). Explicit listeners, on the other hand, like the ones in Figure~\ref{fig:events}, are meant to be used by centralized programs which do not extend class \cmd{Node}.

Listing~\ref{algo:recorder} gives one such example, consisting of a mobility trace recorder. This program listens to topological events of various kinds, including appearance or disappearance of nodes or links, and movements of the nodes. Upon each of these events, it outputs a string representation of the event using a dedicated human readable format called DGS~\cite{DGOP07}. Similar code could be written for Gephi~\cite{gephi}.

\begin{algorithm}[h]
\begin{lstlisting}[language=Java, firstline=8]
public class MyRecorder implements TopologyListener,
                               ConnectivityListener,
                               MovementListener {
   public MyRecorder(Topology tp){
      tp.addTopologyListener(this);
      tp.addConnectivityListener(this);
      tp.addMovementListener(this);
   }

   // TopologyListener
   public void onNodeAdded(Node node) {
      System.out.println("an " + node.hashCode() + 
              " x:" + node.getX() + " y:" + node.getY());
   }


   public void onNodeRemoved(Node node) {
      System.out.println("dn " + node.hashCode());
   }

   // ConnectivityListener
   public void onLinkAdded(Link link) {
      System.out.println("ae " + link.hashCode() + " " + 
              link.endpoint(0) + " "  + link.endpoint(1));
   }
   public void onLinkRemoved(Link link) {
      System.out.println("de " + link.hashCode());
   }

   // MovementListener
   public void onMovement(Node node) {
      System.out.println("cn " + node.hashCode() + 
              " x:" + node.getX() + " y:" + node.getY());
   }
}
\end{lstlisting}
\caption{\label{algo:recorder}Example of a mobility trace recorder}
\end{algorithm}

Other events exist besides those represented in Figure~\ref{fig:events}, such as the \cmd{SelectionListener} interface, which makes it possible to be notified when a node is selected (middle-click) and make it initiate some tasks, for instance broadcast, distinguished role, etc.

\newpage

\subsection{Single threading: why and how?}

It seems convenient at first, to assign every node a dedicated thread, however \jbotsim was designed differently.  \jbotsim is single-threaded, and definitely so. This section explains the why and the how. Understanding these aspects are instrumental in developing well-organized and bug-free programs. 

In \jbotsim, all the nodes, and in fact all of \jbotsim's life (GUI excepted) is articulated around a single thread, which is driven by the central {\em clock}. The clock pulses at regular interval (whose period can be tuned) and notifies its listeners in a specific order. \jbotsim's internal engines, such as the message engine, are served first. Then come those nodes whose wait period has expired (remind that nodes can choose to register to the clock with different periods). These nodes are notified in a random order. Hence, if all nodes listen to the clock at a rate of $1$ (the default value), they will all be notified in a random order in each round, which makes \jbotsim's scheduler a non-deterministic 2-bounded fair scheduler. (Other policies will be available eventually.) A simplified version of the current scheduling process is depicted on Figure~\ref{fig:clock}.

\tikzset{titled/.style={draw, rectangle, rectangle split, rectangle split parts=2}}
\tikzset{normal/.style={draw, rectangle}}
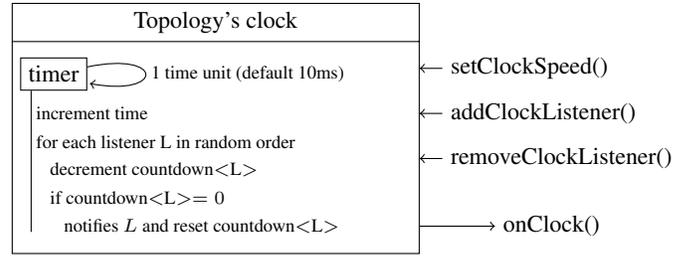
\begin{figure}[h]
  \begin{tikzpicture}
  \path (0,0) node[titled] (clock){
    Topology's clock
    \nodepart{second}
    \begin{minipage}[l]{5.2cm}
      \begin{tikzpicture}[yscale=.7]
        \path (0,0) node[normal] (timer){timer};
        \path (timer.south)+(-.3,0) coordinate (timerr);
        \tikzstyle{every node}=[font=\scriptsize,inner sep=0]
        \draw (timer)+(-.3,-3) -- node[pos=.8,right] {\ increment time}
                                node[pos=.6,right] {\ for each listener L in random order}
                                node[pos=.4,right] {\ ~~~decrement countdown$<$L$>$} 
                                node[pos=.2,right] {\ ~~~if countdown$<$L$>=0$} 
                                node[pos=.0,right] {\ ~~~~~~notifies $L$ and reset countdown$<$L$>$}
                                (timerr);
        \path (timer) edge[loop right] node[xshift=2pt,yshift=-2pt]{\scriptsize 1 time unit (default 10ms)}(timer);
      \end{tikzpicture}
    \end{minipage}
  };
  \tikzstyle{every node}=[font=\footnotesize]
  \path[->] ([yshift=.8cm]clock.east)+(.3,0) node[right] {setTimeUnit()} edge ([yshift=.8cm]clock.east);
  \path[->] ([yshift=.2cm]clock.east)+(.3,0) node[right] {addClockListener()} edge ([yshift=.2cm]clock.east);
  \path[->] ([yshift=-.4cm]clock.east)+(.3,0) node[right] {removeClockListener()} edge ([yshift=-.4cm]clock.east);
  \path[<-] ([yshift=-1.3cm]clock.east)+(1,0) node[right] {onClock()} edge ([yshift=-1.3cm]clock.east);
\end{tikzpicture}
  \caption{\label{fig:clock} Simplified version of the internal scheduler.}
\end{figure}

One consequence of single-threading is that all computations (GUI excepted) take place in a sequential order that makes it possible to use unsynchronized data structures and simpler code. This also improves the scalability of \jbotsim when the number of nodes grows large. One can rely on other user-defined threads in the program, however one should be careful that these thread do not interfer with \jbotsim's. The canonical example is when a scenario is set up by program from within the thread of the \cmd{main()} method. If the initialization makes extensive use of \jbotsim's API from within that thread and the clock starts triggering events at the same time, then problems might occur (and a \cmd{Concurrent\allowbreak ModificationException} be raised). The easy way around is to pause the clock before executing these instructions and to resume it after (using the \cmd{pause()} and the \cmd{resume()} methods on the topology, respectively).

\subsection{Interactivity}
\label{sec:interactivity}

\jbotsim was designed with a clear separation in mind between GUI and internal features. In particular, it can be run without GUI (i.e. without creating the \cmd{JViewer} object), and things will work exactly the same, though invisibly. As such, \jbotsim can be used to perform batch simulations (e.g. sequences of unattended runs that log the effects of some varying parameter). This also enables to withstand heavier simulations in terms of the number of nodes and links.

This being said, one of the most distinctive features of \jbotsim remains {\em interactivity}, e.g., the ability to challenge the algorithm in difficult configurations through adding, removing, or moving nodes during the execution. This approach proves useful to think of a problem visually and intuitively. It also makes it possible to explain someone an algorithm through showing its behavior.

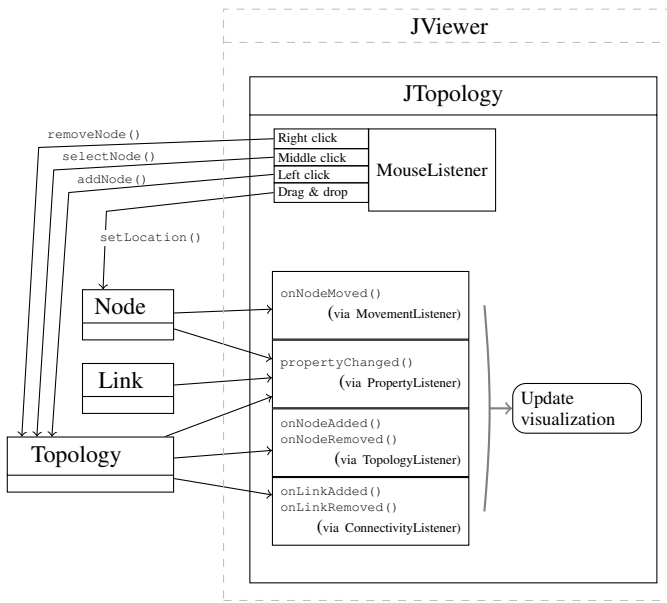
\begin{figure}[h]
\tikzset{titled/.style={draw, rectangle, font=\normalsize, rectangle split, rectangle split parts=2}}
\tikzset{normal/.style={draw, rectangle, font=\scriptsize}}
\tikzset{button/.style={normal, font=\tiny, text width=1.15cm, inner sep=1.475pt, minimum height=0cm}}
\tikzset{listener/.style={normal, text width=2.4cm,minimum height=.9cm}}
\begin{tikzpicture}
  \path (0,0) node[titled,above right] (jtopology){
    JTopology
    \nodepart{second}
    \begin{minipage}[c]{5.2cm}
      \vspace{6cm}~
    \end{minipage}
  };

  \path (1.58,5.5) node[normal,right,minimum height=1.093cm] (mouseL){MouseListener};
  \path (mouseL.north west) node[button,xshift=.3pt,below left] (mright){Right click};
  \path (mright.south east) node[button,yshift=.3pt,below left] (mmiddle){Middle click};
  \path (mmiddle.south east) node[button,yshift=.3pt,below left] (mleft){Left click};
  \path (mleft.south west) node[button,yshift=.4pt,below right] (mdrag){Drag \& drop};

  \path (0.3,3.7) node[listener,right] (movementL){{\tt \tiny onMovement()}\\
    \hfill (\tiny via MovementListener)};
  \path (movementL.south west) node[listener,below right] (propertyL){{\tt \tiny onPropertyChanged()}\\
    \hfill (\tiny via PropertyListener)};
  \path (propertyL.south west) node[listener,below right] (topologyL){{\tt \tiny onNodeAdded()~~~\vspace{-2pt}\\
    onNodeRemoved()}\\\hfill (\tiny via TopologyListener)};
  \path (topologyL.south west) node[listener,below right] (connectivityL){{\tt \tiny onLinkAdded()~~~\vspace{-2pt}\\
    onLinkRemoved()}\\\hfill (\tiny via ConnectivityListener)};

  \draw ([xshift=.2cm]movementL.east) edge[thick,gray,bend left=5] coordinate[midway] (vizbase)([xshift=.2cm]connectivityL.east);
  \draw (vizbase) node[normal,right,xshift=.3cm,text width=1.5cm,rounded corners=5pt]{Update\\visualization} edge[<-,thick,gray] (vizbase);

  \path (jtopology.west)+(-1,.2) node[titled,text width=1cm,left] (node){\,Node\nodepart{second}\vspace{-10pt}};
  \path (node.south east)+(0,-.3) node[titled,text width=1cm,below left] (link){~Link\nodepart{second}\vspace{-10pt}};
  \path (link.south east)+(0,-.3) node[titled,text width=2cm,below left] (topology){~~Topology\nodepart{second}\vspace{-10pt}};

  \tikzstyle{every node}=[font=\scriptsize]
  \draw (movementL) edge[<-] (node);
  \draw (propertyL.west)+(0,.2) edge[<-] (node);
  \draw (propertyL.west)+(0,-.05) edge[<-] (link);
  \draw (propertyL.west)+(0,-.3) edge[<-] (topology);
  \draw (topologyL) edge[<-] (topology);
  \draw (connectivityL) edge[<-] (topology);

  \tikzstyle{every node}=[font=\tt\tiny]
  \path (topology.north west)+(.2,0) coordinate (topoNW1);
  \path (topology.north west)+(.4,0) coordinate (topoNW2);
  \path (topology.north west)+(.6,0) coordinate (topoNW3);
  \path (node.north west)+(.28,0) coordinate (nodeNW);
  \draw[->] (mdrag.west) -- (-1.9,4.95) -- (nodeNW);
  \path (-1.3,4.6) node[fill=white,inner sep=1pt] {setLocation()};
  \draw[->] (mleft.west) -- (-2.4,5.2) node[above right]{addNode()} -- (topoNW3);
  \draw[->] (mmiddle.west) -- (-2.6,5.5) node[above right]{selectNode()} -- (topoNW2);
  \draw[->] (mright.west) -- (-2.8,5.8) node[above right]{removeNode()} -- (topoNW1);

  \path (jtopology.north west)+(-.35,.9) node[titled,dashed,lightgray,below right] {{\color{black}JViewer}
    \nodepart{second}
    \begin{minipage}[c]{5.8cm}
      \vspace{7.2cm}~
    \end{minipage}    
  };
\end{tikzpicture}
\caption{\label{fig:viewer}Internals of JBotSim's GUI}
\end{figure}

The architecture of \jbotsim's viewer is depicted on Figure~\ref{fig:viewer}. 
As one can see, the viewer relies heavily on events related to nodes, links, and topology. The influence also goes the other way, with mouse actions being translated into topological operations. These features are realized by a class called \cmd{JTopology}. This class can often be ignored by the developer, which creates and manipulates the viewer through the higher \cmd{JViewer} class. The latter adds external features such as tuning slide bars, popup menus, or self-containment in a system window.

While natural to \jbotsim's users, the viewer remains, in all technical aspects, an independent piece of software. Alternative viewers could very well be designed with specific uses in mind. 


\section{A zoom on selected features}
\label{sec:zoom}
This section provides more details on a handful of selected features. It covers, namely, the topics of message passing and graph-level algorithms.

\subsection{Exchanging messages}

The way messages are used in \jbotsim is independent from the communication technology considered. Indeed, the API is quite simple, messages are sent directly by the sender, through calling the \cmd{send()} method on its own instance (inherited from \cmd{Node}). Messages are typically received through overriding the \cmd{onMessage()} method (also from class \cmd{Node}). Another way to receive messages, which is not event-based, is for a node to check its mailbox manually through the \cmd{getMailbox()} method, for instance when it is executing the \cmd{onClock()} method (this is the natural way to implement round-based communication models).

Listing~\ref{algo:message} shows a message-based implementation of the flooding principle. Initially, none of the nodes are informed. Then, if a node is selected (either through middle-click or through direct call to the \cmd{selectNode()} method), then this node is notified in the \cmd{onSelection()} method and initiates a basic broadcast scheme. Here, the algorithm consists in retransmitting the received message upon first reception to all the local neighbors (\cmd{sendAll()}).
\begin{algorithm}
\begin{lstlisting}[language=Java, firstline=4]
public class FloodingNode extends Node {

    boolean informed = false;

    @Override
    public void onSelection() {
        informed = true;
        sendAll(new Message());
    }

    @Override
    public void onMessage(Message message) {
        if (!informed){
            informed = true;
            sendAll(message);
        }
    }
}
\end{lstlisting}
\caption{\label{algo:message} Example of message passing algorithm}
\end{algorithm}
In this example, the message is empty, however in general any object can be inserted in the message (no copy is made and the very same object is to be delivered, unless a copy is made at sending time). By default, a message takes one time unit to be transmitted (i.e. it is delivered at the next clock pulse). The delivery happens only for those links which exist by the time of reception. Since any object can be used as message content, it often has to be cast upon reception. All these aspects correspond to \jbotsim's default message engine. Other message engines can be written, and indeed some exist in the \cmd{io.jbotsim.contrib.messaging} package.

\subsection{Working at the graph level}

Implementing an algorithm in the message passing model is sometimes difficult. \jbotsim makes it possible to sketch an idea, play with it, and share it with others, in minutes, thanks to working at a (more abstract) graph-level. This level of abstraction also is relevant in its own right, when the object of study is itself at the graph level, such as classical {\em graph algorithms} or distributed coarse-grain models like {\em graph relabeling systems}~\cite{grs} or {\em population protocols}~\cite{pp}.

To illustrate the simplicity of the graph level, let us consider a scenario where a type of node called \cmd{SocialNode}, dislikes being isolated. Such a node is happy (green) if it has at least one neighbor, unhappy (red) otherwise. In the message passing paradigm, this principle would require to send periodic messages (beacons) and track the reception of these messages, as well as using a timer to decide when a node becomes isolated. Listing~\ref{algo:socialnode} shows a possible implementation at the graph level, which is pretty concise and self-explanatory. 
\begin{algorithm}[h]
\begin{lstlisting}[language=Java, firstline=5]
public class SocialNode extends Node {
    public SocialNode(){
        setColor(Color.RED);
    }
    @Override
    public void onLinkAdded(Link l){
        setColor(Color.GREEN);
    }
    @Override
    public void onLinkRemoved(Link l){
        if (!hasNeighbors())
            setColor(Color.RED);
    }
}
\end{lstlisting}
\caption{\label{algo:socialnode} Example of graph-based algorithm}
\end{algorithm}
Topological events are directly detected by the nodes, which can update their status. (These events could also be listened to globally through the \cmd{ConnectivityListener} interface. This is the way one might want to implement centralized dynamic graph algorithms.) 
Of course, from a message passing perspective, this implementation is cheating. Thus, methods like \cmd{onLinkAdded()}, \cmd{hasNeighbors()}, or \cmd{getNeighbors()} should not be used in a message-passing setting.

\section{Concluding remarks}
\label{sec:extensions}

\jbotsim is mainly a {\em kernel} (\cmd{jbotsim-core} on {\it Maven}), in the sense that it encapsulates a number of generic features whose purpose is to be used by higher programs. 
As of today, one of our goals is to try and keep this kernel simple. 
This does not mean, of course, that \jbotsim\ should not be extended {\em externally}. 

\begin{figure}[h]
  \subfigure[Geocasting in sensor networks]{\input{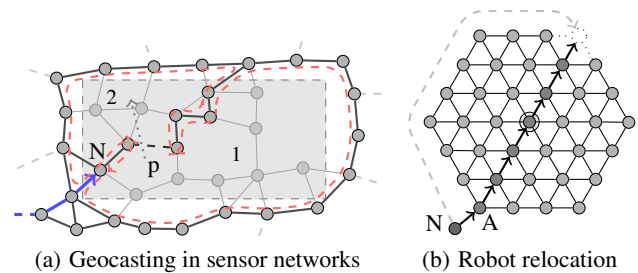}}
  \subfigure[Robot relocation]{    \begin{tikzpicture}[xscale=.55,yscale=.55]
      \tikzstyle{every node}=[draw,circle,inner sep=1.6,fill=black!30]
      \path (2.48,9.34) node (v4) {};
      \path (3.28,9.34) node (v11) {};
      \path (2.88,8.64) node (v7) {};
      \path (3.68,10.03) node (v15) {};
      \path (2.88,10.03) node (v8) {};
      \path (3.28,10.72) node (v12) {};
      \path (4.08,10.72) node (v19) {};
      \path (5.28,10.03) node (v29) {};
      \path (4.88,9.34) node (v25) {};
      \path (4.48,8.64) node (v21) {};
      \path (5.68,9.34) node (v31) {};
      \path (5.28,8.64) node (v28) {};
      \path (4.88,7.95) node (v24) {};
      \path (4.08,7.95) node (v17) {};
      \path (5.68,10.72) node (v32) {};
      \path (4.48,11.42) node (v23) {};
      \path (3.68,11.42) node (v16) {};
      \path (2.88,11.42) node (v9) {};
      \path (2.48,10.72) node (v5) {};
      \path (2.08,10.03) node (v2) {};
      \path (6.08,10.03) node (v34) {};
      \path (6.48,9.33) node (v35) {};
      \path (6.08,8.64) node (v33) {};
      \path (5.68,7.95) node (v30) {};
      \path (5.28,7.26) node (v27) {};
      \path (4.48,7.26) node (v20) {};
      \path (3.68,7.26) node (v13) {};
      \path (2.48,7.95) node (v3) {};
      \path (2.08,8.65) node (v1) {};
      \path (1.68,9.34) node (v0) {};

      \tikzstyle{every node}=[draw,circle,inner sep=2.6,fill=gray!20]
      \path (4.08,9.34) node (poi) {};

      \tikzstyle{every node}=[draw,circle,inner sep=1.6,fill=black!50]
      \path (4.08,9.34) node (v18) {};
      \path (3.68,8.64) node (v14) {};
      \path (4.48,10.03) node (v22) {};
      \path (3.28,7.95) node (v10) {};
      \path (4.88,10.72) node (v26) {};
      \path (2.88,7.26) node (v6) {};

      \tikzstyle{every path}=[];
      \draw (v4)--(v11);
      \draw (v4)--(v7);
      \draw (v11)--(v7);
      \draw (v7)--(v14);
      \draw (v11)--(v14);
      \draw (v11)--(v15);
      \draw (v11)--(v8);
      \draw (v15)--(v8);
      \draw (v15)--(v22);
      \draw (v14)--(v18);
      \draw (v15)--(v18);
      \draw (v22)--(v18);
      \draw (v4)--(v8);
      \draw (v18)--(v11);
      \draw (v22)--(v19);
      \draw (v19)--(v15);
      \draw (v8)--(v12);
      \draw (v19)--(v12);
      \draw (v15)--(v12);
      \draw (v22)--(v26);
      \draw (v19)--(v26);
      \draw (v22)--(v29);
      \draw (v26)--(v29);
      \draw (v22)--(v25);
      \draw (v18)--(v25);
      \draw (v29)--(v25);
      \draw (v14)--(v21);
      \draw (v18)--(v21);
      \draw (v25)--(v21);
      \draw (v10)--(v7);
      \draw (v14)--(v10);
      \draw (v29)--(v31);
      \draw (v25)--(v31);
      \draw (v25)--(v28);
      \draw (v21)--(v28);
      \draw (v31)--(v28);
      \draw (v21)--(v24);
      \draw (v28)--(v24);
      \draw (v14)--(v17);
      \draw (v10)--(v17);
      \draw (v21)--(v17);
      \draw (v24)--(v17);
      \draw (v26)--(v32);
      \draw (v29)--(v32);
      \draw (v19)--(v23);
      \draw (v26)--(v23);
      \draw (v12)--(v16);
      \draw (v19)--(v16);
      \draw (v23)--(v16);
      \draw (v12)--(v9);
      \draw (v16)--(v9);
      \draw (v8)--(v5);
      \draw (v9)--(v5);
      \draw (v12)--(v5);
      \draw (v4)--(v2);
      \draw (v8)--(v2);
      \draw (v5)--(v2);
      \draw (v29)--(v34);
      \draw (v31)--(v34);
      \draw (v32)--(v34);
      \draw (v31)--(v35);
      \draw (v34)--(v35);
      \draw (v31)--(v33);
      \draw (v28)--(v33);
      \draw (v35)--(v33);
      \draw (v28)--(v30);
      \draw (v24)--(v30);
      \draw (v33)--(v30);
      \draw (v24)--(v27);
      \draw (v30)--(v27);
      \draw (v24)--(v20);
      \draw (v17)--(v20);
      \draw (v27)--(v20);
      \draw (v10)--(v13);
      \draw (v17)--(v13);
      \draw (v20)--(v13);
      \draw (v10)--(v6);
      \draw (v13)--(v6);
      \draw (v7)--(v3);
      \draw (v10)--(v3);
      \draw (v6)--(v3);
      \draw (v4)--(v1);
      \draw (v7)--(v1);
      \draw (v3)--(v1);
      \draw (v4)--(v0);
      \draw (v2)--(v0);
      \draw (v1)--(v0);

      \tikzstyle{every node}=[draw,circle,inner sep=1.6, fill=black!60]
      \path (2.28,6.76) node (vnew) {};
      \tikzstyle{every node}=[]
      \path[] (vnew)+(-.5,.12) node (labelN) {\footnotesize N};
      \path[] (v6)+(.25,-.35) node (labelA) {\footnotesize A};

      \tikzstyle{every node}=[draw,circle,inner sep=2,dotted]
      \path (5.28,11.42) node (vempty) {};

      \tikzstyle{every path}=[dashed, dotted];
      \draw (v23)--(vempty);
      \draw (v32)--(vempty);

      \tikzstyle{every path}=[->, thick];
      \draw (vnew)--(v6);
      \draw (v6)--(v10);
      \draw (v10)--(v14);
      \draw (v14)--(v18);
      \draw (v18)--(v22);
      \draw (v22)--(v26);
      \draw (v26)--(vempty);

      \tikzstyle{every path}=[->,rounded corners=.2cm, thick, dashed, gray!50];
      \path (v0)+(-.6,.0) coordinate (anc1);
      \path (v9)+(-.3,.6) coordinate (anc2);
      \path (vempty)+(-.3,.6) coordinate (anc3);
      \draw (vnew)--(anc1)--(anc2)--(anc3)--(vempty);
    \end{tikzpicture}    }
  \caption{\label{fig:tikz}Two examples of pictures whose underlying topology was generated using \jbotsim. The topology on the left was created by adding and moving nodes using the mouse; the topology on the right was generated by program. Once exported as a Ti{\it k}z picture, they were twicked manually to add shading, color, etc.}
\end{figure}

In particular, \jbotsim distribution  (\cmd{jbotsim-all} on {\it Maven}) already ships a number of not-mandatory-but-yet-useful features. 
For instance, the \cmd{TikzTopologySerializer} makes it possible to export a topology as a Ti{\it k}Z picture -- a powerful format for drawing pictures in \LaTeX\ documents (see Figure~\ref{fig:tikz}).


Other extensions include basic topology algorithms for testing, e.g., if a given topology is connected or 2-connected, if a given node is critical (its removal would disconnect the graph) or compute the diameter. A set of extensions dedicated to dynamic graph are currently being developped. For instance, the \cmd{EMEGPlayer} takes as input a birth rate, death rate, and an underlying graph (given as a \cmd{Topology}), and generates an edge-markovian dynamic graph based on these parameters, the dynamics of which can be listened to through the \cmd{ConnectivityListener} interface.

Contributions are most welcome, as well as suggestions of improvement or feature requests.




\end{document}